\documentclass[twocolumn, pra, showpacs,superscriptaddress]{revtex4-1}
\usepackage{graphicx}
\usepackage{amsmath}
\usepackage{bm}
\usepackage{float}
\usepackage{color}
\usepackage{sidecap}
\allowdisplaybreaks
\usepackage{soul}
\setstcolor{red}
\usepackage{ulem}

\begin{document}
\title{Harmonic trap resonance enhanced synthetic atomic spin-orbit coupling}
\author{Ling-Na Wu}
\affiliation{State Key Laboratory of Low Dimensional Quantum Physics, Department of Physics, Tsinghua University, Beijing 100084, China}
%\affiliation{Collaborative Innovation Center of Quantum Matter, Beijing, China}
\author{Xin-Yu Luo}
\affiliation{State Key Laboratory of Low Dimensional Quantum Physics, Department of Physics, Tsinghua University, Beijing 100084, China}
%\affiliation{Collaborative Innovation Center of Quantum Matter, Beijing, China}
\author{Zhi-Fang Xu}
\affiliation{Department of Physics, University of Tokyo, 7-3-1 Hongo, Bunkyo-ku, Tokyo 113-0033, Japan}
\affiliation{MOE Key Laboratory of Fundamental Physical Quantities Measurements,
School of Physics, Huazhong University of Science and Technology, Wuhan 430074, China}
\author{Masahito Ueda}
\affiliation{Department of Physics, University of Tokyo, 7-3-1 Hongo, Bunkyo-ku, Tokyo 113-0033, Japan}
\author{Ruquan Wang}
\email{ruquanwang@aphy.iphy.ac.cn}
\affiliation{Institute of Physics, Chinese Academy of Sciences, Beijing 100080,
People¡¯s Republic of China}
\affiliation{Collaborative Innovation Center of Quantum Matter, Beijing, China}
\author{L. You}
\email{lyou@mail.tsinghua.edu.cn}
\affiliation{State Key Laboratory of Low Dimensional Quantum Physics, Department of Physics, Tsinghua University, Beijing 100084, China}
\affiliation{Collaborative Innovation Center of Quantum Matter, Beijing, China}

\date{\today}

\begin{abstract}
Spin-orbit coupling (SOC) plays an essential role in many exotic and interesting phenomena
in condensed matter physics. In neutral-atom-based quantum simulations,
synthetic SOC constitutes a key enabling element.
The strength of SOC realized so far is limited by various reasons or constraints.
This work reports tunable SOC synthesized with a gradient magnetic field (GMF) for atoms in a harmonic trap.
Nearly ten-fold enhancement is observed when the GMF is
modulated near the harmonic-trap resonance in comparison with free-space atoms.
A theory is developed that well explains the experimental results. Our work
offers a clear physical insight into and analytical understanding of how to tune the strength of atomic SOC synthesized
with GMF using harmonic trap resonance.

\end{abstract}

% 67.85.De	Dynamic properties of condensates; excitations, and superfluid flow

%\pacs{67.85.Jk, 03.75.Mn, 03.75.Ss, 37.10.Jk}
%\pacs{03.75.Mn, 67.85.Fg, 67.85.Jk}
%\pacs{67.85.Jk, 03.75.Mn, 03.75.Ss, 37.10.Jk}
%\pacs{03.75.Mn, 03.75.Hh, 67.85.Fg}
%03.75.Hh   Static properties of condensates; thermodynamical, statistical, and structural properties
%67.85.Jk   Other Bose-Einstein condensation phenomena
%03.75.Kk, 03.75.Mn, 67.85.Fg
%05.30.Jp   Boson systems (for static and dynamic properties
%of Bose-Einstein condensates, see 03.75.Hh and 03.75.Kk; see also 67.10.Ba Boson degeneracy in quantum fluids)
%67.85.Fg   Multicomponent condensates; spinor condensates
%05.30.Jp   Boson systems
%67.85.-d   Ultracold gases, trapped gases (see also 03.75.-b Matter waves in quantum mechanics)
%03.75.Mn   Multicomponent condensates; spinor condensates
%03.75.Kk   Dynamic properties of condensates, collective and hydrodynamic excitations, superfluid flow
%03.75.Ss 	Degenerate Fermi gases
%37.10.Jk 	Atoms in optical lattices
%67.60.Bc   Boson mixtures
%67.86.De   Dynamic properties of condensates, excitations, and superfluid flow
%05.45.Xt   Synchronization; coupled oscillators
%05.45.Pq   Numerical simulations of chaotic systems

\maketitle

%\section{Introduction}

Resonance phenomena \cite{Landau1976} frequently occur in Nature. When the frequency of
a time-periodic external drive matches a system's resonance,
the response is dramatic. A folklore wisdom warns that
soldiers crossing a bridge should not march in unison
to prevent its collapsing from
accidentally stepping onto the resonance.
Even at multiple resonant frequencies, such as parametric resonance when the driving frequency is twice the system's characteristic frequency,
the response can still be quite substantial.
In quantum mechanics, resonances become ubiquitous as a result of
quantization, where stationary states of a system feature
definite eigenenergies. A transition between two eigenstates
is resonantly enhanced when the frequency of an external coupling matches
their energy difference \cite{rabi1938,foot2004atomic}.

This work reports our experimental observation and theoretical vindication of an enhanced atomic spin-orbit coupling (SOC)
synthesized with a modulating gradient magnetic field (GMF) applied to atoms in a harmonic trap.
SOC, which couples a particle's spin to its orbital motion,
constitutes one of the most important interactions in condensed matter physics.
In the strong coupling regime, SOC gives rise to nontrivial topological bands,
which support many exotic states and phenomena, including topological band/Mott insulator,
quantum-number fractionalization and magneto-electric effects \cite{pesin2010mott,witczak2014correlated}.
In recent years, atomic quantum gases
have emerged as powerful quantum simulators for condensed matter systems \cite{bloch2012quantum,lewenstein2007ultracold}.
Strong atomic SOC
often plays crucial roles in the increasing list of desired ingredients for artificial gauge fields
~\cite{Dalibard2011,Galitski2013,Cooper2008,Fetter2009,Lin2009b,Lin2011, Aidelsburger2013,Miyake2013,aidelsburger2015measuring,Struck19082011,Struck2012,Parker2013,Jotzu2014}.

The breakthrough on synthetic SOC
came in 2011 ~\cite{Lin2011}, when Spielman's group observed a special type of one-dimensional (1D)
SOC: an equal-weighted sum of Rashba~\cite{RashbaJPC1984} and Dresselhaus~\cite{DresselhausPR1955} types of SOC
created by the momentum-sensitive Raman coupling between two internal states of $^{87}$Rb atoms.
Since then, the Raman scheme has become the prototype for studies involving 1D atomic SOC \cite{PanPRL2012,*PanNat2014,ChenPRA2014,EngelsPRL2015,Cheuk2012,ZhangPRL2012,*ZhangNat2014}.
Recently, the observations of two-dimensional (2D) SOC relying on atom-photon interactions have been reported~\cite{ZhangNatPhys2016,2015arXiv151108492M,2015arXiv151108170W}.
In the Raman scheme, the strength of synthesized SOC is limited by
photon momentum transfer and constrained by Raman laser beam geometry.
A protocol for tuning the strength of SOC including switching its sign through periodically
modulating effective Rabi frequency \cite{Zhang2013} in the Raman scheme has also been realized~\cite{PhysRevLett.114.125301}.

An alternative method of creating synthetic atomic SOC is to use pulsed
or time-periodic GMF ~\cite{Xu2013,Anderson2013,Goldman2014,Struck2014,LuoSR2016,JotzuPRL2015},
which can be implemented free from atomic spontaneous emission.
Its underlying mechanism is the Stern-Gerlach effect,
whereby the periodic GMF
imparts a spin-dependent momentum impulse to the atomic center-of-mass motion.
This spin-dependent impulse can be described in terms of the same linear coupling
between the spin (or pseudo-spin) with the atomic orbital (center-of-mass) motion
as in the Raman scheme.
Through concatenating GMF pulses along two orthogonal directions,
genuine Rashba, Dresselhaus, or even arbitrary types of SOC in 2D can be synthesized for
atoms with arbitrary hyperfine spins \cite{Xu2013,Anderson2013,Goldman2014}.
Very recently, essential features demonstrating 1D tunable SOC
synthesized with a periodic modulating GMF have been reported~\cite{LuoSR2016}.

\begin{figure*}[!htp]
\includegraphics[width=2\columnwidth]{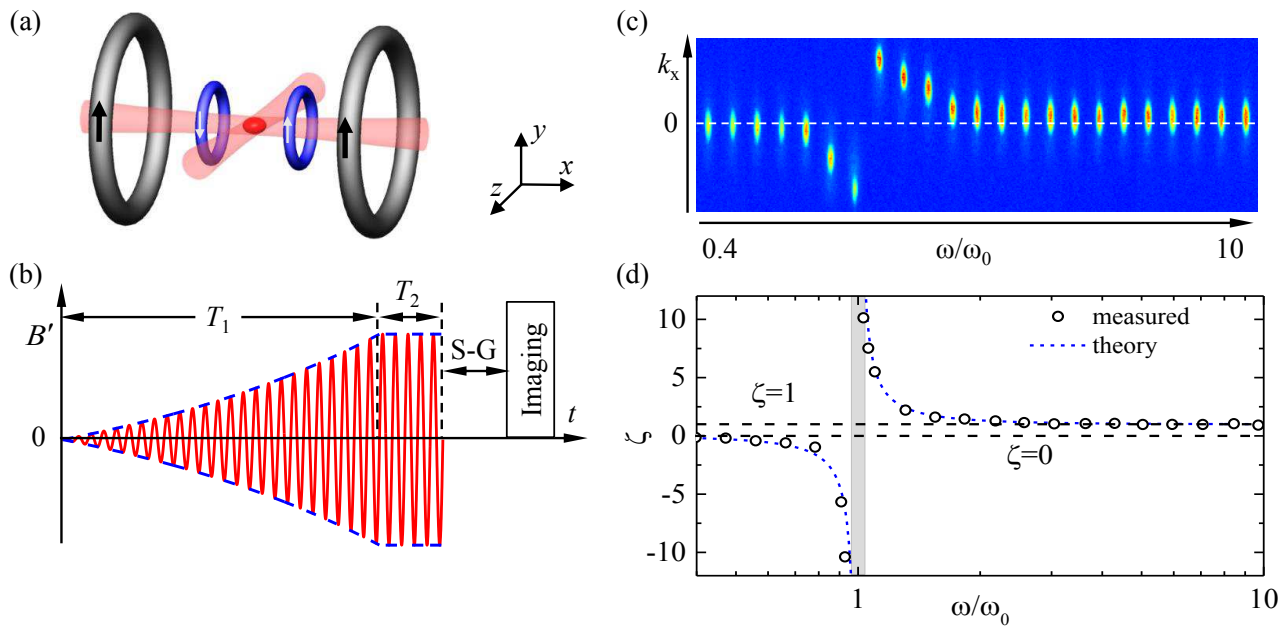}\\
\caption{(Color online)
Harmonic-trap-resonance enhanced SOC. (a) Schematic illustration of the experimental setup, consisting of the bias (gray) and gradient (blue) magnetic
coils. The condensate (red football shape) is produced at the center of a crossed optical dipole trap formed from laser beams in pink.
Its location coincides with the center of the gradient coil
configuration. (b) Time sequence of our experiments. The modulation amplitude (delimited by the blue dashed envelop) of the gradient magnetic field $B'(t)$ (shown in red) is
adiabatically ramped up to an effective value corresponding to $k_{\rm so}=1.25\, \mu m^{-1}$
within $T_1$ and held on for $T_2$, followed by the Stern-Gerlach (S-G) separation
before absorption imaging. To ensure adiabaticity during the ramp, $T_1=250$ms and $T_2=50$ms are chosen for the modulation frequency $\omega> (2\pi)\, 100$ Hz and $T_1=25\tau$ and $T_2=5\tau$, with $\tau=2\pi/\omega$ being the modulation period, for $\omega< (2\pi)\, 100$ Hz. (c) Absorption images for
the momentum-shifted atomic clouds in the $|m_F=\!-1\rangle$ state at different values of $\omega$.
Darker red denotes higher optical density.
The abscissa is not to scale.
Each measured off-set atomic cloud corresponds to a data point
shown in (d) in the same order of increasing modulation frequency from left to right.
In the shaded band region in (d) surrounding the trap resonance, the driven atomic cloud is unable to reach a stationary state yet after $T_2$.
The dashed line denotes $k_x=0$ for without GMF or SOC.
(d) The measured values of the scaled SOC strength $\zeta$ (black open circles)
as a function of $\omega$,
which agree perfectly with Eq. (\ref{cs}) shown
in the blue dotted curve.
} \label{fig1}
\end{figure*}

This work presents a different method for controlling the strength of atomic SOC synthesized with GMF
by making use of harmonic trap resonance for atomic center-of-mass motion. It is beyond the straightforward scheme of tuning
the strength of momentum impulse as demonstrated recently~\cite{LuoSR2016}.
Furthermore, it differs from the reported tuning scheme \cite{PhysRevLett.114.125301}
based on the amplitude-modulated Raman coupling \cite{Zhang2013},
which can only decrease the strength of SOC. The harmonic trap resonance scheme we report here
opens a different avenue for reaching the strong SOC regime.

Our experiment is inspired by the success of synthesizing atomic SOC from
a time-periodic GMF \cite{Anderson2013,LuoSR2016}.
Specifically,
under a periodically modulating 1D GMF, atomic center-of-mass experiences a spin-dependent force, whose overall effect is
simply to shift the momentum from $p_x$ to $p_x-m_F \hbar k_x^{({\rm{min}})}$ for the atomic spin component $m_F$,
where $\hbar k_x^{({\rm{min}})}$ denotes the minimum of the modified dispersion curve.
The momentum of the equilibrium state is thus translated by a spin-dependent amount $m_F \hbar k_x^{({\rm{min}})}$,
which is equivalent to a synthetic SOC with strength $\hbar k_x^{({\rm{min}})}$.
Experimentally, this effective strength of SOC is determined from the measured displacement of the atomic cloud.

A typical experiment starts with a $^{87}$Rb condensate of $1.2\times 10^5$
atoms in the state $\left| {F=1,m_F=-1} \right\rangle$ confined inside a crossed dipole trap
whose minimum potential region is approximately harmonic with
frequencies $(\omega_x,\omega_y,\omega_z)=2\pi \times (77,136,77)$ Hz along three orthogonal spatial directions $x$, $y$, and $z$.
The 1D GMF is implemented by a combination of a 3D quadrupole magnetic field $\textbf{B}_q=B'x{\hat x}-B'y{\hat y}/2-B'z{\hat
z}/2$ and a 5.7 Gauss bias field
$\textbf{B}_b=B_0\hat{x}$ [Fig. \ref{fig1}(a)], whose linear and quadratic Zeeman shifts correspond to $(2\pi)\, 4$ MHz and $(2\pi)\, 2.34$ kHz, respectively. More details about the magnetic field control is as described
in Ref. \cite{LuoSR2016}.
The amplitude for the GMF is sinusoidally modulated as $B'(t)=B'_{\rm{max}}\sin\left(\omega t\right)$,
which translates into a 1D SOC strength $\hbar k_{\rm{so}}=g_F \mu_B B'_{\rm{max}}/\omega$, where
$g_F$ denotes the Land\'e g-factor and $\mu_B$ the Bohr magneton, provided
the time-periodic modulation frequency $\omega$ is far away from trap resonance,
as confirmed in a recent experiment \cite{LuoSR2016}.

As shown in Fig.~\ref{fig1}(b), the orbital momentum of the $|m_F=-1\rangle$ condensate is shifted
to a final equilibrium value corresponding to the SOC strength of $k_{\rm so }=1.25 \mu m^{-1}$
by adiabatically ramping up the GMF modulation amplitude
within 250 ms (or 25 modulation periods for $\omega< (2\pi)\, 100$ Hz),
and then held on for another 50 ms (or 5 modulation periods for $\omega<(2\pi)\,100$ Hz).
At integer multiples of the modulation period $\tau=2\pi/\omega$,
the crossed dipole trap is turned off in less than $10$ $\mu$s.
Subsequently, the condensate expands for about $24$ ms,
during which different Zeeman components are Stern-Gerlach separated
by an inhomogeneous magnetic field along the vertical direction.
A bimodal fit to the atomic cloud density profile measured
 through the standard absorption imaging, as shown in Fig. \ref{fig1}(c), yields the shifted center-of-mass position
 for the condensate.
The spatial displacement from that without SOC is used
to derive the momentum shift $k_{x}^{\rm (min)}$,
from which the scaled SOC strength $\zeta=k_{x}^{\rm (min)}/k_{\rm so}$ is computed.

A clear resonance behavior is observed for $\zeta$, as shown in Figs. \ref{fig1}(c) and (d),
for its dependence on the modulation frequency $\omega$ relative to the trap frequency $\omega_0=\omega_x=(2\pi)\,77$Hz.
Above the trap frequency $\omega_0$, $\zeta$ increases with decreasing $\omega$,
from $\zeta=1$ for $\omega$ far above the resonance to a peak
when $\omega$ approaches $\omega_0$.
Below the trap frequency, $\zeta$ changes its sign, with its magnitude growing from $\zeta=0$ for $\omega$ far below the resonance to a peak around the resonance.
The enhanced response on the opposite sides of the resonance
is out of phase as a result of the
$\pi$ phase shift across a resonance.
Limited by our present setup,
we operate in the regime of
small momentum impulse and observe nearly ten-fold enhancement for $\zeta$ at $\omega/\omega_0=1.03$,
where heating remains insignificant. The effect of heating-induced damping becomes noticeable in the immediate vicinity
of resonance.

This dramatic resonance enhancement of SOC due to the harmonic trap cannot be explained by
the previous theory for atoms in free space \cite{Anderson2013},
which neglects the influence of the trapping potential on atomic motion.
One might have naively concluded that an analogous calculation that incorporates the effect of the trap potential into the
previously studied free-space model would find the agreement with the observed resonance.
Unfortunately, this is easily said than done.
To demonstrate it, we briefly recapitulate the basic idea of the previous theory \cite{Anderson2013}.
For an atom of mass $m$ in free space and in the presence of
a sinusoidally modulating GMF
along the $x$-direction, the effective 1D Hamiltonian is given by
\begin{eqnarray}
{H}_0(t) = \frac{p_x^2}{2m} + \beta(t)\hbar k_{\rm so}xF_x,
\label{hfree}
\end{eqnarray}
where $p_x$ is the momentum of an atom,
$F_x$ is the $x$-component of its spin $F$, and $\beta(t)=\omega \sin\left(\omega t\right)$ is
the temporal profile for the coupling strength
between the time-dependent GMF and
atomic magnetic dipole moment measured in units of modulation amplitude
$\hbar k_{\rm so}$.
The Schr\"odinger equation for ${H}_0(t)$ of Eq. (\ref{hfree}) can be more easily
handled if we introduce a unitary transformation $\psi=R(t)\tilde{\psi}$
with $R(t)=\exp[-ix\,A_x(t)/\hbar]$, which corresponds to
 a momentum translation by the spin-dependent impulse
$A_{x}(t)=\hbar k_{\rm so}F_x\int_0^t\beta(t')dt'=\hbar k_{\rm so}[1-\cos(\omega t)]F_x$ from the GMF.
The wave function $\tilde{\psi}$ in the rotating frame is
then governed by the momentum shifted Hamiltonian
$
{\tilde H}_0(t) = \frac{{1}}{{2m}}R^{\dag}(t)p_x^2R(t)
={[ {p_x} - A_x(t) ]}^2/2m,
$
which commutes with itself at different times, $[\tilde H_0(t),\tilde H_0(t')]=0$.
The corresponding time evolution operator takes a simple form
$\tilde U_0(t)=\exp[-i\int_0^{t}\tilde H_0(t_1)dt_1/\hbar]$.
After one period of evolution $\tau=2\pi/\omega$, we obtain $A_x(\tau)=0$, or $R(\tau)=1$.
Hence, the wavefunctions in the two frames coincide, $\psi(\tau)=\tilde{\psi}(\tau)$, and the
effective Hamiltonian for the whole period is given by
\begin{eqnarray} \label{heff0}
{H_{{\rm{eff}}}^{(0)}} &=& \frac{1}{\tau}\int_0^{\tau}  {\tilde {H_0}} (t)dt\ \notag\\
 &=& \frac{{{{\left( {p_x} - \hbar {k_{{\rm{so}}}}{F_x} \right)}^2}}}{{2m}} + \frac{{\hbar ^2}k_{{\rm{so}}}^2}{4m} F_x^2,
\end{eqnarray}
where the first term describes the SOC of strength $\hbar {k_{{\rm{so}}}}$ and the second term acts like a quadratic Zeeman shift.

In the presence of a 1D harmonic trap $V_{\rm trap}=m\omega_0^2x^2/2$, the Hamiltonian changes into
\begin{eqnarray}\label{Htrap}
 H(t)=H_0(t)+\frac{1}{2}m\omega_0^2x^2,
  \label{li2}
\end{eqnarray}
which in the rotating frame becomes
\begin{eqnarray}
\tilde H(t)=\frac{{{{[ {p_x} - A_x(t) ]}^2}}}{{2m}}+\frac{1}{2}m\omega_0^2x^2.
  \label{li2}
\end{eqnarray}
Unlike the case of a free atom discussed above,
the two $\tilde H(t)$'s in Eq. (\ref{li2}) at different times do not
always commute due to the presence of $V_{\rm trap}$.
The corresponding unitary evolution operator then takes a more general form
$\tilde U(t)=\mathcal{T}\exp\left[-i\int_0^{t}\tilde H(t_1)dt_1/\hbar\right]$,
where $\mathcal{T}$ denotes time ordering. This unitary evolution operator takes such
a complicated form that it is difficult to derive the effective Hamiltonian
in a straightforward manner. Therefore, we have to resort to other means for
a compact solution capable of explaining the resonant behavior observed.

We note that Hamiltonian (\ref{Htrap}) also describes a sinusoidally driven harmonic oscillator,
whose quantum-mechanical propagator can be obtained in the explicit analytic form.
Hence, we can get the effective Hamiltonian of the system by making use of the propagator.
For Hamiltonian (\ref{Htrap}), its propagator is given by
\begin{widetext}
\begin{eqnarray}\label{propagatorforce}
  K(x'',\tau;x',0)
  &=&\sqrt{\frac{m\omega_0}{2\pi i\hbar\sin\omega_0\tau}}\exp\left\{
    i\frac{m\omega_0}{2\hbar\sin\omega_0\tau}\Bigg[(x'^2+x''^2)\cos\omega_0\tau-2x'x''\Bigg]\right\}
    \nonumber\\
    &&\times\exp\left[i\frac{1}{1-\omega_0^2/\omega^2}k_{\rm so}(x''-x')F_x
    -i\frac{1}{2-2\omega_0^2/\omega^2}\frac{\hbar k_{\rm so}^2}{2m}F_x^2\tau\right].
\end{eqnarray}
\end{widetext}
The effective Hamiltonian of the system should give
the same propagator as Eq. (\ref{propagatorforce}).
Without loss of generality,
it is reasonable to infer that the effective Hamiltonian
will not be much different from that in free space in Eq.~(\ref{heff0}).
We therefore assume
\begin{eqnarray}\label{Hsoc}
  H_{\rm{eff}}&=&\frac{(p_x-\zeta\hbar k_{\rm so}F_x)^2}{2m}+\frac{1}{2}m\omega_0^2x^2
  +s\frac{\hbar^2k_{\rm so}^2}{2m}F_x^2,
\end{eqnarray}
where $\zeta$ and $s$ denote modifications to
the strength of SOC and the quadratic Zeeman shift, respectively,
when the trap potential $V_{\rm trap}$ is present.
The corresponding propagator in this case is found to be
\begin{widetext}
\begin{eqnarray}
  K(x'',\tau;x',0)
  &=&\sqrt{\frac{m\omega_0}{2\pi i\hbar\sin\omega_0\tau}}\exp\left\{
    i\frac{m\omega_0}{2\hbar\sin\omega_0\tau}\Bigg[(x'^2+x''^2)\cos\omega_0\tau-2x'x''\Bigg]\right\},
    \nonumber\\
  &&\times\exp\left[i\zeta k_{\rm so}(x''-x')F_x-is\frac{\hbar k_{\rm so}^2}{2m}F_x^2\tau\right].
  \label{propagatorsoc}
\end{eqnarray}
\end{widetext}
The equivalence between the two propagators of Eqs. (\ref{propagatorforce}) and (\ref{propagatorsoc})
thus gives
\begin{eqnarray}\label{cs}
\zeta=\frac{1}{1-\omega_0^2/\omega^2},
\end{eqnarray}
and $s=\frac{1}{2}\zeta$. In other words, the effective Hamiltonian for atoms in a harmonic trap driven by a sinusoidally modulating GMF is found to be given by Eq. (\ref{Hsoc}), which is similar in form to
the case of a free atom except for $\zeta$ and $s$ in Eq. (\ref{cs}).

The factor $\zeta$ in Eq. (\ref{cs}) is plotted as the blue dotted curve in
Fig. \ref{fig1}(d), which is found to agree well with the measured data
except in the immediate vicinity of the trap resonance [shaded band region in Fig. \ref{fig1}(d)], where the finite lifetime of the condensate makes it difficult to reach equilibrium.
It clearly shows that as the modulation frequency $\omega$ approaches the trap frequency $\omega_0$,
the SOC becomes enhanced. More specifically, when approaching
the resonance from above, the effective SOC is increasingly enhanced.
Upon crossing the resonance $\omega_0$,
the effective SOC reverses its sign, and the factor $\zeta$ gradually decreases
and eventually tails off to zero at frequencies much smaller than $\omega_0$.
This dependence of $\zeta$ on the modulation frequency highlights the
tunability discussed in this work.
In the immediate vicinity of the resonance,
the amplitude of atomic micro-motion due to periodic modulation is so large that the Gaussian-shaped optical trap cannot be well approximated
by a harmonic trap anymore, and as a result of the large amplitude oscillations the condensate collapses.

For the 1D situation considered here,
the associated gauge field can be transformed away because the
corresponding 1D SOC describes a simple spin-dependent momentum shift.
When tracking the dynamics for the different atomic spin components,
the accumulated phase from the SOC term, however, is real as confirmed in the recent experiment \cite{LuoSR2016}.
Furthermore, in the presence of a uniform bias magnetic field,
which gives rise to an interaction $\propto F_z$,
or when any other non-commuting interactions are included, the SOC discussed
above persists and cannot be gauged away even for the 1D system~\cite{LuoSR2016,You2016PhysRevLett.116.143003}.

In conclusion, for the atomic SOC synthesized from a time-periodic GMF,
we observed a resonant behavior which highlights nearly ten-fold enhanced
SOC when the modulation frequency is close to but higher than the trap frequency.
This resonance is accompanied by
a progression towards vanishing SOC on the lower modulation frequency side and a reduction to the
value for a free atom in the higher modulation frequency side.
We develop a theory that well explains the experimentally observed resonant behavior.
Compared with atoms in free space under a sinusoidally modulating GMF, we find
that an effective SOC Hamiltonian for atoms confined inside a harmonic trap
takes an analogous form, except for a frequency-dependent prefactor.
This prefactor reveals the
resonant behaviour as the periodic drive hits the motional resonance of the harmonic trap.

We thank Qifeng Xie for the assistance of the experiment and Jinlong Yu for helpful discussions.
This work is supported by MOST 2013CB922002
and 2013CB922004 of the National Key Basic Research
Program of China, and by NSFC (No. 91121005,
No. 11404184, No. 11574100, and No. 11474347).

%\bibliography{reference}{}

\begin{thebibliography}{41}%
\makeatletter
\providecommand \@ifxundefined [1]{%
 \@ifx{#1\undefined}
}%
\providecommand \@ifnum [1]{%
 \ifnum #1\expandafter \@firstoftwo
 \else \expandafter \@secondoftwo
 \fi
}%
\providecommand \@ifx [1]{%
 \ifx #1\expandafter \@firstoftwo
 \else \expandafter \@secondoftwo
 \fi
}%
\providecommand \natexlab [1]{#1}%
\providecommand \enquote  [1]{``#1''}%
\providecommand \bibnamefont  [1]{#1}%
\providecommand \bibfnamefont [1]{#1}%
\providecommand \citenamefont [1]{#1}%
\providecommand \href@noop [0]{\@secondoftwo}%
\providecommand \href [0]{\begingroup \@sanitize@url \@href}%
\providecommand \@href[1]{\@@startlink{#1}\@@href}%
\providecommand \@@href[1]{\endgroup#1\@@endlink}%
\providecommand \@sanitize@url [0]{\catcode `\\12\catcode `\$12\catcode
  `\&12\catcode `\#12\catcode `\^12\catcode `\_12\catcode `\%12\relax}%
\providecommand \@@startlink[1]{}%
\providecommand \@@endlink[0]{}%
\providecommand \url  [0]{\begingroup\@sanitize@url \@url }%
\providecommand \@url [1]{\endgroup\@href {#1}{\urlprefix }}%
\providecommand \urlprefix  [0]{URL }%
\providecommand \Eprint [0]{\href }%
\providecommand \doibase [0]{http://dx.doi.org/}%
\providecommand \selectlanguage [0]{\@gobble}%
\providecommand \bibinfo  [0]{\@secondoftwo}%
\providecommand \bibfield  [0]{\@secondoftwo}%
\providecommand \translation [1]{[#1]}%
\providecommand \BibitemOpen [0]{}%
\providecommand \bibitemStop [0]{}%
\providecommand \bibitemNoStop [0]{.\EOS\space}%
\providecommand \EOS [0]{\spacefactor3000\relax}%
\providecommand \BibitemShut  [1]{\csname bibitem#1\endcsname}%
\let\auto@bib@innerbib\@empty
%</preamble>
\bibitem [{\citenamefont {{Landau}}\ and\ \citenamefont
  {{Lifshitz}}(1976)}]{Landau1976}%
  \BibitemOpen
  \bibfield  {author} {\bibinfo {author} {\bibfnamefont {L.~D.}\ \bibnamefont
  {{Landau}}}\ and\ \bibinfo {author} {\bibfnamefont {E.~M.}\ \bibnamefont
  {{Lifshitz}}},\ }\href
  {http://gen.lib.rus.ec/book/index.php?md5=02C2029CBB3BEF0EDD817193C5FDEB0E}
  {\emph {\bibinfo {title} {Mechanics}}},\ \bibinfo {edition} {3rd}\ ed.,\
  \bibinfo {series} {Course of Theoretical Physics}, Vol.~\bibinfo {volume}
  {1}\ (\bibinfo  {publisher} {Butterworth-Heinemann},\ \bibinfo {year}
  {1976})\BibitemShut {NoStop}%
\bibitem [{\citenamefont {{Rabi}}\ \emph {et~al.}(1938)\citenamefont {{Rabi}},
  \citenamefont {{Zacharias}}, \citenamefont {{Millman}},\ and\ \citenamefont
  {{Kusch}}}]{rabi1938}%
  \BibitemOpen
  \bibfield  {author} {\bibinfo {author} {\bibfnamefont {I.~I.}\ \bibnamefont
  {{Rabi}}}, \bibinfo {author} {\bibfnamefont {J.~R.}\ \bibnamefont
  {{Zacharias}}}, \bibinfo {author} {\bibfnamefont {S.}~\bibnamefont
  {{Millman}}}, \ and\ \bibinfo {author} {\bibfnamefont {P.}~\bibnamefont
  {{Kusch}}},\ }\href {\doibase 10.1103/PhysRev.53.318} {\bibfield  {journal}
  {\bibinfo  {journal} {Physical Review}\ }\textbf {\bibinfo {volume} {53}},\
  \bibinfo {pages} {318} (\bibinfo {year} {1938})}\BibitemShut {NoStop}%
\bibitem [{\citenamefont {Foot}(2004)}]{foot2004atomic}%
  \BibitemOpen
  \bibfield  {author} {\bibinfo {author} {\bibfnamefont {C.~J.}\ \bibnamefont
  {Foot}},\ }\href@noop {} {\emph {\bibinfo {title} {Atomic Physics}}}\
  (\bibinfo  {publisher} {Oxford University Press, Oxford},\ \bibinfo {year}
  {2004})\BibitemShut {NoStop}%
\bibitem [{\citenamefont {Pesin}\ and\ \citenamefont
  {Balents}(2010)}]{pesin2010mott}%
  \BibitemOpen
  \bibfield  {author} {\bibinfo {author} {\bibfnamefont {D.}~\bibnamefont
  {Pesin}}\ and\ \bibinfo {author} {\bibfnamefont {L.}~\bibnamefont
  {Balents}},\ }\href@noop {} {\bibfield  {journal} {\bibinfo  {journal}
  {Nature Physics}\ }\textbf {\bibinfo {volume} {6}},\ \bibinfo {pages} {376}
  (\bibinfo {year} {2010})}\BibitemShut {NoStop}%
\bibitem [{\citenamefont {Witczak-Krempa}\ \emph {et~al.}(2014)\citenamefont
  {Witczak-Krempa}, \citenamefont {Chen}, \citenamefont {Kim},\ and\
  \citenamefont {Balents}}]{witczak2014correlated}%
  \BibitemOpen
  \bibfield  {author} {\bibinfo {author} {\bibfnamefont {W.}~\bibnamefont
  {Witczak-Krempa}}, \bibinfo {author} {\bibfnamefont {G.}~\bibnamefont
  {Chen}}, \bibinfo {author} {\bibfnamefont {Y.~B.}\ \bibnamefont {Kim}}, \
  and\ \bibinfo {author} {\bibfnamefont {L.}~\bibnamefont {Balents}},\ }\href
  {\doibase 10.1146/annurev-conmatphys-020911-125138} {\bibfield  {journal}
  {\bibinfo  {journal} {Annual Review of Condensed Matter Physics}\ }\textbf
  {\bibinfo {volume} {5}},\ \bibinfo {pages} {57} (\bibinfo {year}
  {2014})}\BibitemShut {NoStop}%
\bibitem [{\citenamefont {Bloch}\ \emph {et~al.}(2012)\citenamefont {Bloch},
  \citenamefont {Dalibard},\ and\ \citenamefont
  {Nascimb{\`e}ne}}]{bloch2012quantum}%
  \BibitemOpen
  \bibfield  {author} {\bibinfo {author} {\bibfnamefont {I.}~\bibnamefont
  {Bloch}}, \bibinfo {author} {\bibfnamefont {J.}~\bibnamefont {Dalibard}}, \
  and\ \bibinfo {author} {\bibfnamefont {S.}~\bibnamefont {Nascimb{\`e}ne}},\
  }\href@noop {} {\bibfield  {journal} {\bibinfo  {journal} {Nature Physics}\
  }\textbf {\bibinfo {volume} {8}},\ \bibinfo {pages} {267} (\bibinfo {year}
  {2012})}\BibitemShut {NoStop}%
\bibitem [{\citenamefont {Lewenstein}\ \emph {et~al.}(2007)\citenamefont
  {Lewenstein}, \citenamefont {Sanpera}, \citenamefont {Ahufinger},
  \citenamefont {Damski}, \citenamefont {Sen},\ and\ \citenamefont
  {Sen}}]{lewenstein2007ultracold}%
  \BibitemOpen
  \bibfield  {author} {\bibinfo {author} {\bibfnamefont {M.}~\bibnamefont
  {Lewenstein}}, \bibinfo {author} {\bibfnamefont {A.}~\bibnamefont {Sanpera}},
  \bibinfo {author} {\bibfnamefont {V.}~\bibnamefont {Ahufinger}}, \bibinfo
  {author} {\bibfnamefont {B.}~\bibnamefont {Damski}}, \bibinfo {author}
  {\bibfnamefont {A.}~\bibnamefont {Sen}}, \ and\ \bibinfo {author}
  {\bibfnamefont {U.}~\bibnamefont {Sen}},\ }\href@noop {} {\bibfield
  {journal} {\bibinfo  {journal} {Advances in Physics}\ }\textbf {\bibinfo
  {volume} {56}},\ \bibinfo {pages} {243} (\bibinfo {year} {2007})}\BibitemShut
  {NoStop}%
\bibitem [{\citenamefont {Dalibard}\ \emph {et~al.}(2011)\citenamefont
  {Dalibard}, \citenamefont {Gerbier}, \citenamefont
  {Juzeli\ifmmode~\bar{u}\else \={u}\fi{}nas},\ and\ \citenamefont
  {\"Ohberg}}]{Dalibard2011}%
  \BibitemOpen
  \bibfield  {author} {\bibinfo {author} {\bibfnamefont {J.}~\bibnamefont
  {Dalibard}}, \bibinfo {author} {\bibfnamefont {F.}~\bibnamefont {Gerbier}},
  \bibinfo {author} {\bibfnamefont {G.}~\bibnamefont
  {Juzeli\ifmmode~\bar{u}\else \={u}\fi{}nas}}, \ and\ \bibinfo {author}
  {\bibfnamefont {P.}~\bibnamefont {\"Ohberg}},\ }\href {\doibase
  10.1103/RevModPhys.83.1523} {\bibfield  {journal} {\bibinfo  {journal} {Rev.
  Mod. Phys.}\ }\textbf {\bibinfo {volume} {83}},\ \bibinfo {pages} {1523}
  (\bibinfo {year} {2011})}\BibitemShut {NoStop}%
\bibitem [{\citenamefont {Galitski}\ and\ \citenamefont
  {Spielman}(2013)}]{Galitski2013}%
  \BibitemOpen
  \bibfield  {author} {\bibinfo {author} {\bibfnamefont {V.}~\bibnamefont
  {Galitski}}\ and\ \bibinfo {author} {\bibfnamefont {I.~B.}\ \bibnamefont
  {Spielman}},\ }\href {\doibase 10.1038/nature11841} {\bibfield  {journal}
  {\bibinfo  {journal} {Nature}\ }\textbf {\bibinfo {volume} {494}},\ \bibinfo
  {pages} {49} (\bibinfo {year} {2013})}\BibitemShut {NoStop}%
\bibitem [{\citenamefont {Cooper}(2008)}]{Cooper2008}%
  \BibitemOpen
  \bibfield  {author} {\bibinfo {author} {\bibfnamefont {N.}~\bibnamefont
  {Cooper}},\ }\href {\doibase 10.1080/00018730802564122} {\bibfield  {journal}
  {\bibinfo  {journal} {Advances in Physics}\ }\textbf {\bibinfo {volume}
  {57}},\ \bibinfo {pages} {539} (\bibinfo {year} {2008})}\BibitemShut
  {NoStop}%
\bibitem [{\citenamefont {Fetter}(2009)}]{Fetter2009}%
  \BibitemOpen
  \bibfield  {author} {\bibinfo {author} {\bibfnamefont {A.~L.}\ \bibnamefont
  {Fetter}},\ }\href {\doibase 10.1103/RevModPhys.81.647} {\bibfield  {journal}
  {\bibinfo  {journal} {Rev. Mod. Phys.}\ }\textbf {\bibinfo {volume} {81}},\
  \bibinfo {pages} {647} (\bibinfo {year} {2009})}\BibitemShut {NoStop}%
\bibitem [{\citenamefont {Lin}\ \emph {et~al.}(2009)\citenamefont {Lin},
  \citenamefont {Compton}, \citenamefont {Jim\'enez-Garc\'{\i}a}, \citenamefont
  {Porto},\ and\ \citenamefont {Spielman}}]{Lin2009b}%
  \BibitemOpen
  \bibfield  {author} {\bibinfo {author} {\bibfnamefont {Y.-J.}\ \bibnamefont
  {Lin}}, \bibinfo {author} {\bibfnamefont {R.~L.}\ \bibnamefont {Compton}},
  \bibinfo {author} {\bibfnamefont {K.}~\bibnamefont {Jim\'enez-Garc\'{\i}a}},
  \bibinfo {author} {\bibfnamefont {J.~V.}\ \bibnamefont {Porto}}, \ and\
  \bibinfo {author} {\bibfnamefont {I.~B.}\ \bibnamefont {Spielman}},\
  }\href@noop {} {\bibfield  {journal} {\bibinfo  {journal} {Nature}\ }\textbf
  {\bibinfo {volume} {462}},\ \bibinfo {pages} {628} (\bibinfo {year}
  {2009})}\BibitemShut {NoStop}%
\bibitem [{\citenamefont {Lin}\ \emph {et~al.}(2011)\citenamefont {Lin},
  \citenamefont {Jim\'enez-Garc\'{\i}a},\ and\ \citenamefont
  {Spielman}}]{Lin2011}%
  \BibitemOpen
  \bibfield  {author} {\bibinfo {author} {\bibfnamefont {Y.-J.}\ \bibnamefont
  {Lin}}, \bibinfo {author} {\bibfnamefont {K.}~\bibnamefont
  {Jim\'enez-Garc\'{\i}a}}, \ and\ \bibinfo {author} {\bibfnamefont {I.~B.}\
  \bibnamefont {Spielman}},\ }\href@noop {} {\bibfield  {journal} {\bibinfo
  {journal} {Nature}\ }\textbf {\bibinfo {volume} {471}},\ \bibinfo {pages}
  {83} (\bibinfo {year} {2011})}\BibitemShut {NoStop}%
\bibitem [{\citenamefont {Aidelsburger}\ \emph {et~al.}(2013)\citenamefont
  {Aidelsburger}, \citenamefont {Atala}, \citenamefont {Lohse}, \citenamefont
  {Barreiro}, \citenamefont {Paredes},\ and\ \citenamefont
  {Bloch}}]{Aidelsburger2013}%
  \BibitemOpen
  \bibfield  {author} {\bibinfo {author} {\bibfnamefont {M.}~\bibnamefont
  {Aidelsburger}}, \bibinfo {author} {\bibfnamefont {M.}~\bibnamefont {Atala}},
  \bibinfo {author} {\bibfnamefont {M.}~\bibnamefont {Lohse}}, \bibinfo
  {author} {\bibfnamefont {J.~T.}\ \bibnamefont {Barreiro}}, \bibinfo {author}
  {\bibfnamefont {B.}~\bibnamefont {Paredes}}, \ and\ \bibinfo {author}
  {\bibfnamefont {I.}~\bibnamefont {Bloch}},\ }\href {\doibase
  10.1103/PhysRevLett.111.185301} {\bibfield  {journal} {\bibinfo  {journal}
  {Phys. Rev. Lett.}\ }\textbf {\bibinfo {volume} {111}},\ \bibinfo {pages}
  {185301} (\bibinfo {year} {2013})}\BibitemShut {NoStop}%
\bibitem [{\citenamefont {Miyake}\ \emph {et~al.}(2013)\citenamefont {Miyake},
  \citenamefont {Siviloglou}, \citenamefont {Kennedy}, \citenamefont {Burton},\
  and\ \citenamefont {Ketterle}}]{Miyake2013}%
  \BibitemOpen
  \bibfield  {author} {\bibinfo {author} {\bibfnamefont {H.}~\bibnamefont
  {Miyake}}, \bibinfo {author} {\bibfnamefont {G.~A.}\ \bibnamefont
  {Siviloglou}}, \bibinfo {author} {\bibfnamefont {C.~J.}\ \bibnamefont
  {Kennedy}}, \bibinfo {author} {\bibfnamefont {W.~C.}\ \bibnamefont {Burton}},
  \ and\ \bibinfo {author} {\bibfnamefont {W.}~\bibnamefont {Ketterle}},\
  }\href {\doibase 10.1103/PhysRevLett.111.185302} {\bibfield  {journal}
  {\bibinfo  {journal} {Phys. Rev. Lett.}\ }\textbf {\bibinfo {volume} {111}},\
  \bibinfo {pages} {185302} (\bibinfo {year} {2013})}\BibitemShut {NoStop}%
\bibitem [{\citenamefont {Aidelsburger}\ \emph {et~al.}(2015)\citenamefont
  {Aidelsburger}, \citenamefont {Lohse}, \citenamefont {Schweizer},
  \citenamefont {Atala}, \citenamefont {Barreiro}, \citenamefont {Nascimbene},
  \citenamefont {Cooper}, \citenamefont {Bloch},\ and\ \citenamefont
  {Goldman}}]{aidelsburger2015measuring}%
  \BibitemOpen
  \bibfield  {author} {\bibinfo {author} {\bibfnamefont {M.}~\bibnamefont
  {Aidelsburger}}, \bibinfo {author} {\bibfnamefont {M.}~\bibnamefont {Lohse}},
  \bibinfo {author} {\bibfnamefont {C.}~\bibnamefont {Schweizer}}, \bibinfo
  {author} {\bibfnamefont {M.}~\bibnamefont {Atala}}, \bibinfo {author}
  {\bibfnamefont {J.~T.}\ \bibnamefont {Barreiro}}, \bibinfo {author}
  {\bibfnamefont {S.}~\bibnamefont {Nascimbene}}, \bibinfo {author}
  {\bibfnamefont {N.}~\bibnamefont {Cooper}}, \bibinfo {author} {\bibfnamefont
  {I.}~\bibnamefont {Bloch}}, \ and\ \bibinfo {author} {\bibfnamefont
  {N.}~\bibnamefont {Goldman}},\ }\href@noop {} {\bibfield  {journal} {\bibinfo
   {journal} {Nature Physics}\ }\textbf {\bibinfo {volume} {11}},\ \bibinfo
  {pages} {162} (\bibinfo {year} {2015})}\BibitemShut {NoStop}%
\bibitem [{\citenamefont {Struck}\ \emph {et~al.}(2011)\citenamefont {Struck},
  \citenamefont {Ölschläger}, \citenamefont {Le~Targat}, \citenamefont
  {Soltan-Panahi}, \citenamefont {Eckardt}, \citenamefont {Lewenstein},
  \citenamefont {Windpassinger},\ and\ \citenamefont
  {Sengstock}}]{Struck19082011}%
  \BibitemOpen
  \bibfield  {author} {\bibinfo {author} {\bibfnamefont {J.}~\bibnamefont
  {Struck}}, \bibinfo {author} {\bibfnamefont {C.}~\bibnamefont
  {Ölschläger}}, \bibinfo {author} {\bibfnamefont {R.}~\bibnamefont
  {Le~Targat}}, \bibinfo {author} {\bibfnamefont {P.}~\bibnamefont
  {Soltan-Panahi}}, \bibinfo {author} {\bibfnamefont {A.}~\bibnamefont
  {Eckardt}}, \bibinfo {author} {\bibfnamefont {M.}~\bibnamefont {Lewenstein}},
  \bibinfo {author} {\bibfnamefont {P.}~\bibnamefont {Windpassinger}}, \ and\
  \bibinfo {author} {\bibfnamefont {K.}~\bibnamefont {Sengstock}},\ }\href
  {\doibase 10.1126/science.1207239} {\bibfield  {journal} {\bibinfo  {journal}
  {Science}\ }\textbf {\bibinfo {volume} {333}},\ \bibinfo {pages} {996}
  (\bibinfo {year} {2011})}\BibitemShut {NoStop}%
\bibitem [{\citenamefont {Struck}\ \emph {et~al.}(2012)\citenamefont {Struck},
  \citenamefont {\"Olschl\"ager}, \citenamefont {Weinberg}, \citenamefont
  {Hauke}, \citenamefont {Simonet}, \citenamefont {Eckardt}, \citenamefont
  {Lewenstein}, \citenamefont {Sengstock},\ and\ \citenamefont
  {Windpassinger}}]{Struck2012}%
  \BibitemOpen
  \bibfield  {author} {\bibinfo {author} {\bibfnamefont {J.}~\bibnamefont
  {Struck}}, \bibinfo {author} {\bibfnamefont {C.}~\bibnamefont
  {\"Olschl\"ager}}, \bibinfo {author} {\bibfnamefont {M.}~\bibnamefont
  {Weinberg}}, \bibinfo {author} {\bibfnamefont {P.}~\bibnamefont {Hauke}},
  \bibinfo {author} {\bibfnamefont {J.}~\bibnamefont {Simonet}}, \bibinfo
  {author} {\bibfnamefont {A.}~\bibnamefont {Eckardt}}, \bibinfo {author}
  {\bibfnamefont {M.}~\bibnamefont {Lewenstein}}, \bibinfo {author}
  {\bibfnamefont {K.}~\bibnamefont {Sengstock}}, \ and\ \bibinfo {author}
  {\bibfnamefont {P.}~\bibnamefont {Windpassinger}},\ }\href {\doibase
  10.1103/PhysRevLett.108.225304} {\bibfield  {journal} {\bibinfo  {journal}
  {Phys. Rev. Lett.}\ }\textbf {\bibinfo {volume} {108}},\ \bibinfo {pages}
  {225304} (\bibinfo {year} {2012})}\BibitemShut {NoStop}%
\bibitem [{\citenamefont {Parker}\ \emph {et~al.}(2013)\citenamefont {Parker},
  \citenamefont {Ha},\ and\ \citenamefont {Chin}}]{Parker2013}%
  \BibitemOpen
  \bibfield  {author} {\bibinfo {author} {\bibfnamefont {C.~V.}\ \bibnamefont
  {Parker}}, \bibinfo {author} {\bibfnamefont {L.-C.}\ \bibnamefont {Ha}}, \
  and\ \bibinfo {author} {\bibfnamefont {C.}~\bibnamefont {Chin}},\ }\href@noop
  {} {\bibfield  {journal} {\bibinfo  {journal} {Nature Physics}\ }\textbf
  {\bibinfo {volume} {9}},\ \bibinfo {pages} {769} (\bibinfo {year}
  {2013})}\BibitemShut {NoStop}%
\bibitem [{\citenamefont {Jotzu}\ \emph {et~al.}(2014)\citenamefont {Jotzu},
  \citenamefont {Messer}, \citenamefont {Desbuquois}, \citenamefont {Lebrat},
  \citenamefont {Uehlinger}, \citenamefont {Greif},\ and\ \citenamefont
  {Esslinger}}]{Jotzu2014}%
  \BibitemOpen
  \bibfield  {author} {\bibinfo {author} {\bibfnamefont {G.}~\bibnamefont
  {Jotzu}}, \bibinfo {author} {\bibfnamefont {M.}~\bibnamefont {Messer}},
  \bibinfo {author} {\bibfnamefont {R.}~\bibnamefont {Desbuquois}}, \bibinfo
  {author} {\bibfnamefont {M.}~\bibnamefont {Lebrat}}, \bibinfo {author}
  {\bibfnamefont {T.}~\bibnamefont {Uehlinger}}, \bibinfo {author}
  {\bibfnamefont {D.}~\bibnamefont {Greif}}, \ and\ \bibinfo {author}
  {\bibfnamefont {T.}~\bibnamefont {Esslinger}},\ }\href@noop {} {\bibfield
  {journal} {\bibinfo  {journal} {Nature}\ }\textbf {\bibinfo {volume} {515}},\
  \bibinfo {pages} {237} (\bibinfo {year} {2014})}\BibitemShut {NoStop}%
\bibitem [{\citenamefont {Bychkov}\ and\ \citenamefont
  {Rashba}(1984)}]{RashbaJPC1984}%
  \BibitemOpen
  \bibfield  {author} {\bibinfo {author} {\bibfnamefont {Y.~A.}\ \bibnamefont
  {Bychkov}}\ and\ \bibinfo {author} {\bibfnamefont {E.~I.}\ \bibnamefont
  {Rashba}},\ }\href@noop {} {\bibfield  {journal} {\bibinfo  {journal} {J.
  Phys. C}\ }\textbf {\bibinfo {volume} {17}},\ \bibinfo {pages} {6039}
  (\bibinfo {year} {1984})}\BibitemShut {NoStop}%
\bibitem [{\citenamefont {Dresselhaus}(1955)}]{DresselhausPR1955}%
  \BibitemOpen
  \bibfield  {author} {\bibinfo {author} {\bibfnamefont {G.}~\bibnamefont
  {Dresselhaus}},\ }\href {\doibase 10.1103/PhysRev.100.580} {\bibfield
  {journal} {\bibinfo  {journal} {Phys. Rev.}\ }\textbf {\bibinfo {volume}
  {100}},\ \bibinfo {pages} {580} (\bibinfo {year} {1955})}\BibitemShut
  {NoStop}%
\bibitem [{\citenamefont {Zhang}\ \emph {et~al.}(2012)\citenamefont {Zhang},
  \citenamefont {Ji}, \citenamefont {Chen}, \citenamefont {Zhang},
  \citenamefont {Du}, \citenamefont {Yan}, \citenamefont {Pan}, \citenamefont
  {Zhao}, \citenamefont {Deng}, \citenamefont {Zhai}, \citenamefont {Chen},\
  and\ \citenamefont {Pan}}]{PanPRL2012}%
  \BibitemOpen
  \bibfield  {author} {\bibinfo {author} {\bibfnamefont {J.-Y.}\ \bibnamefont
  {Zhang}}, \bibinfo {author} {\bibfnamefont {S.-C.}\ \bibnamefont {Ji}},
  \bibinfo {author} {\bibfnamefont {Z.}~\bibnamefont {Chen}}, \bibinfo {author}
  {\bibfnamefont {L.}~\bibnamefont {Zhang}}, \bibinfo {author} {\bibfnamefont
  {Z.-D.}\ \bibnamefont {Du}}, \bibinfo {author} {\bibfnamefont
  {B.}~\bibnamefont {Yan}}, \bibinfo {author} {\bibfnamefont {G.-S.}\
  \bibnamefont {Pan}}, \bibinfo {author} {\bibfnamefont {B.}~\bibnamefont
  {Zhao}}, \bibinfo {author} {\bibfnamefont {Y.-J.}\ \bibnamefont {Deng}},
  \bibinfo {author} {\bibfnamefont {H.}~\bibnamefont {Zhai}}, \bibinfo {author}
  {\bibfnamefont {S.}~\bibnamefont {Chen}}, \ and\ \bibinfo {author}
  {\bibfnamefont {J.-W.}\ \bibnamefont {Pan}},\ }\href {\doibase
  10.1103/PhysRevLett.109.115301} {\bibfield  {journal} {\bibinfo  {journal}
  {Phys. Rev. Lett.}\ }\textbf {\bibinfo {volume} {109}},\ \bibinfo {pages}
  {115301} (\bibinfo {year} {2012})}\BibitemShut {NoStop}%
\bibitem [{\citenamefont {Ji}\ \emph {et~al.}(2014)\citenamefont {Ji},
  \citenamefont {Zhang}, \citenamefont {Zhang}, \citenamefont {Du},
  \citenamefont {Zheng}, \citenamefont {Deng}, \citenamefont {Zhai},
  \citenamefont {Chen},\ and\ \citenamefont {Pan}}]{PanNat2014}%
  \BibitemOpen
  \bibfield  {author} {\bibinfo {author} {\bibfnamefont {S.-C.}\ \bibnamefont
  {Ji}}, \bibinfo {author} {\bibfnamefont {J.-Y.}\ \bibnamefont {Zhang}},
  \bibinfo {author} {\bibfnamefont {L.}~\bibnamefont {Zhang}}, \bibinfo
  {author} {\bibfnamefont {Z.-D.}\ \bibnamefont {Du}}, \bibinfo {author}
  {\bibfnamefont {W.}~\bibnamefont {Zheng}}, \bibinfo {author} {\bibfnamefont
  {Y.-J.}\ \bibnamefont {Deng}}, \bibinfo {author} {\bibfnamefont
  {H.}~\bibnamefont {Zhai}}, \bibinfo {author} {\bibfnamefont {S.}~\bibnamefont
  {Chen}}, \ and\ \bibinfo {author} {\bibfnamefont {J.-W.}\ \bibnamefont
  {Pan}},\ }\href@noop {} {\bibfield  {journal} {\bibinfo  {journal} {Nature
  Physics}\ }\textbf {\bibinfo {volume} {10}},\ \bibinfo {pages} {314}
  (\bibinfo {year} {2014})}\BibitemShut {NoStop}%
\bibitem [{\citenamefont {Olson}\ \emph {et~al.}(2014)\citenamefont {Olson},
  \citenamefont {Wang}, \citenamefont {Niffenegger}, \citenamefont {Li},
  \citenamefont {Greene},\ and\ \citenamefont {Chen}}]{ChenPRA2014}%
  \BibitemOpen
  \bibfield  {author} {\bibinfo {author} {\bibfnamefont {A.~J.}\ \bibnamefont
  {Olson}}, \bibinfo {author} {\bibfnamefont {S.-J.}\ \bibnamefont {Wang}},
  \bibinfo {author} {\bibfnamefont {R.~J.}\ \bibnamefont {Niffenegger}},
  \bibinfo {author} {\bibfnamefont {C.-H.}\ \bibnamefont {Li}}, \bibinfo
  {author} {\bibfnamefont {C.~H.}\ \bibnamefont {Greene}}, \ and\ \bibinfo
  {author} {\bibfnamefont {Y.~P.}\ \bibnamefont {Chen}},\ }\href {\doibase
  10.1103/PhysRevA.90.013616} {\bibfield  {journal} {\bibinfo  {journal} {Phys.
  Rev. A}\ }\textbf {\bibinfo {volume} {90}},\ \bibinfo {pages} {013616}
  (\bibinfo {year} {2014})}\BibitemShut {NoStop}%
\bibitem [{\citenamefont {Hamner}\ \emph {et~al.}(2015)\citenamefont {Hamner},
  \citenamefont {Zhang}, \citenamefont {Khamehchi}, \citenamefont {Davis},\
  and\ \citenamefont {Engels}}]{EngelsPRL2015}%
  \BibitemOpen
  \bibfield  {author} {\bibinfo {author} {\bibfnamefont {C.}~\bibnamefont
  {Hamner}}, \bibinfo {author} {\bibfnamefont {Y.}~\bibnamefont {Zhang}},
  \bibinfo {author} {\bibfnamefont {M.~A.}\ \bibnamefont {Khamehchi}}, \bibinfo
  {author} {\bibfnamefont {M.~J.}\ \bibnamefont {Davis}}, \ and\ \bibinfo
  {author} {\bibfnamefont {P.}~\bibnamefont {Engels}},\ }\href {\doibase
  10.1103/PhysRevLett.114.070401} {\bibfield  {journal} {\bibinfo  {journal}
  {Phys. Rev. Lett.}\ }\textbf {\bibinfo {volume} {114}},\ \bibinfo {pages}
  {070401} (\bibinfo {year} {2015})}\BibitemShut {NoStop}%
\bibitem [{\citenamefont {Cheuk}\ \emph {et~al.}(2012)\citenamefont {Cheuk},
  \citenamefont {Sommer}, \citenamefont {Hadzibabic}, \citenamefont {Yefsah},
  \citenamefont {Bakr},\ and\ \citenamefont {Zwierlein}}]{Cheuk2012}%
  \BibitemOpen
  \bibfield  {author} {\bibinfo {author} {\bibfnamefont {L.~W.}\ \bibnamefont
  {Cheuk}}, \bibinfo {author} {\bibfnamefont {A.~T.}\ \bibnamefont {Sommer}},
  \bibinfo {author} {\bibfnamefont {Z.}~\bibnamefont {Hadzibabic}}, \bibinfo
  {author} {\bibfnamefont {T.}~\bibnamefont {Yefsah}}, \bibinfo {author}
  {\bibfnamefont {W.~S.}\ \bibnamefont {Bakr}}, \ and\ \bibinfo {author}
  {\bibfnamefont {M.~W.}\ \bibnamefont {Zwierlein}},\ }\href {\doibase
  10.1103/PhysRevLett.109.095302} {\bibfield  {journal} {\bibinfo  {journal}
  {Phys. Rev. Lett.}\ }\textbf {\bibinfo {volume} {109}},\ \bibinfo {pages}
  {095302} (\bibinfo {year} {2012})}\BibitemShut {NoStop}%
\bibitem [{\citenamefont {Wang}\ \emph {et~al.}(2012)\citenamefont {Wang},
  \citenamefont {Yu}, \citenamefont {Fu}, \citenamefont {Miao}, \citenamefont
  {Huang}, \citenamefont {Chai}, \citenamefont {Zhai},\ and\ \citenamefont
  {Zhang}}]{ZhangPRL2012}%
  \BibitemOpen
  \bibfield  {author} {\bibinfo {author} {\bibfnamefont {P.}~\bibnamefont
  {Wang}}, \bibinfo {author} {\bibfnamefont {Z.-Q.}\ \bibnamefont {Yu}},
  \bibinfo {author} {\bibfnamefont {Z.}~\bibnamefont {Fu}}, \bibinfo {author}
  {\bibfnamefont {J.}~\bibnamefont {Miao}}, \bibinfo {author} {\bibfnamefont
  {L.}~\bibnamefont {Huang}}, \bibinfo {author} {\bibfnamefont
  {S.}~\bibnamefont {Chai}}, \bibinfo {author} {\bibfnamefont {H.}~\bibnamefont
  {Zhai}}, \ and\ \bibinfo {author} {\bibfnamefont {J.}~\bibnamefont {Zhang}},\
  }\href {\doibase 10.1103/PhysRevLett.109.095301} {\bibfield  {journal}
  {\bibinfo  {journal} {Phys. Rev. Lett.}\ }\textbf {\bibinfo {volume} {109}},\
  \bibinfo {pages} {095301} (\bibinfo {year} {2012})}\BibitemShut {NoStop}%
\bibitem [{\citenamefont {Fu}\ \emph {et~al.}(2014)\citenamefont {Fu},
  \citenamefont {Huang}, \citenamefont {Meng}, \citenamefont {Wang},
  \citenamefont {Zhang}, \citenamefont {Zhang}, \citenamefont {Zhai},
  \citenamefont {Zhang},\ and\ \citenamefont {Zhang}}]{ZhangNat2014}%
  \BibitemOpen
  \bibfield  {author} {\bibinfo {author} {\bibfnamefont {Z.}~\bibnamefont
  {Fu}}, \bibinfo {author} {\bibfnamefont {L.}~\bibnamefont {Huang}}, \bibinfo
  {author} {\bibfnamefont {Z.}~\bibnamefont {Meng}}, \bibinfo {author}
  {\bibfnamefont {P.}~\bibnamefont {Wang}}, \bibinfo {author} {\bibfnamefont
  {L.}~\bibnamefont {Zhang}}, \bibinfo {author} {\bibfnamefont
  {S.}~\bibnamefont {Zhang}}, \bibinfo {author} {\bibfnamefont
  {H.}~\bibnamefont {Zhai}}, \bibinfo {author} {\bibfnamefont {P.}~\bibnamefont
  {Zhang}}, \ and\ \bibinfo {author} {\bibfnamefont {J.}~\bibnamefont
  {Zhang}},\ }\href@noop {} {\bibfield  {journal} {\bibinfo  {journal} {Nature
  Physics}\ }\textbf {\bibinfo {volume} {10}},\ \bibinfo {pages} {110}
  (\bibinfo {year} {2014})}\BibitemShut {NoStop}%
\bibitem [{\citenamefont {Huang}\ \emph {et~al.}(2015)\citenamefont {Huang},
  \citenamefont {Meng}, \citenamefont {Wang}, \citenamefont {Peng},
  \citenamefont {Zhang}, \citenamefont {Chen}, \citenamefont {Li},
  \citenamefont {Zhou},\ and\ \citenamefont {Zhang}}]{ZhangNatPhys2016}%
  \BibitemOpen
  \bibfield  {author} {\bibinfo {author} {\bibfnamefont {L.}~\bibnamefont
  {Huang}}, \bibinfo {author} {\bibfnamefont {Z.}~\bibnamefont {Meng}},
  \bibinfo {author} {\bibfnamefont {P.}~\bibnamefont {Wang}}, \bibinfo {author}
  {\bibfnamefont {P.}~\bibnamefont {Peng}}, \bibinfo {author} {\bibfnamefont
  {S.-L.}\ \bibnamefont {Zhang}}, \bibinfo {author} {\bibfnamefont
  {L.}~\bibnamefont {Chen}}, \bibinfo {author} {\bibfnamefont {D.}~\bibnamefont
  {Li}}, \bibinfo {author} {\bibfnamefont {Q.}~\bibnamefont {Zhou}}, \ and\
  \bibinfo {author} {\bibfnamefont {J.}~\bibnamefont {Zhang}},\ }\href@noop {}
  {\bibfield  {journal} {\bibinfo  {journal} {Nature Physics}\ }\textbf
  {\bibinfo {volume} {115}},\ \bibinfo {pages} {073002} (\bibinfo {year}
  {2015})}\BibitemShut {NoStop}%
\bibitem [{\citenamefont {{Meng}}\ \emph {et~al.}(2015)\citenamefont {{Meng}},
  \citenamefont {{Huang}}, \citenamefont {{Peng}}, \citenamefont {{Li}},
  \citenamefont {{Chen}}, \citenamefont {{Xu}}, \citenamefont {{Zhang}},
  \citenamefont {{Wang}},\ and\ \citenamefont {{Zhang}}}]{2015arXiv151108492M}%
  \BibitemOpen
  \bibfield  {author} {\bibinfo {author} {\bibfnamefont {Z.}~\bibnamefont
  {{Meng}}}, \bibinfo {author} {\bibfnamefont {L.}~\bibnamefont {{Huang}}},
  \bibinfo {author} {\bibfnamefont {P.}~\bibnamefont {{Peng}}}, \bibinfo
  {author} {\bibfnamefont {D.}~\bibnamefont {{Li}}}, \bibinfo {author}
  {\bibfnamefont {L.}~\bibnamefont {{Chen}}}, \bibinfo {author} {\bibfnamefont
  {Y.}~\bibnamefont {{Xu}}}, \bibinfo {author} {\bibfnamefont {C.}~\bibnamefont
  {{Zhang}}}, \bibinfo {author} {\bibfnamefont {P.}~\bibnamefont {{Wang}}}, \
  and\ \bibinfo {author} {\bibfnamefont {J.}~\bibnamefont {{Zhang}}},\
  }\href@noop {} {\bibfield  {journal} {\bibinfo  {journal} {arXiv:
  1511.08492}\ } (\bibinfo {year} {2015})}\BibitemShut {NoStop}%
\bibitem [{\citenamefont {{Wu}}\ \emph {et~al.}()\citenamefont {{Wu}},
  \citenamefont {{Zhang}}, \citenamefont {{Sun}}, \citenamefont {{Xu}},
  \citenamefont {{Wang}}, \citenamefont {{Ji}}, \citenamefont {{Deng}},
  \citenamefont {{Chen}}, \citenamefont {{Liu}},\ and\ \citenamefont
  {{Pan}}}]{2015arXiv151108170W}%
  \BibitemOpen
  \bibfield  {author} {\bibinfo {author} {\bibfnamefont {Z.}~\bibnamefont
  {{Wu}}}, \bibinfo {author} {\bibfnamefont {L.}~\bibnamefont {{Zhang}}},
  \bibinfo {author} {\bibfnamefont {W.}~\bibnamefont {{Sun}}}, \bibinfo
  {author} {\bibfnamefont {X.-T.}\ \bibnamefont {{Xu}}}, \bibinfo {author}
  {\bibfnamefont {B.-Z.}\ \bibnamefont {{Wang}}}, \bibinfo {author}
  {\bibfnamefont {S.-C.}\ \bibnamefont {{Ji}}}, \bibinfo {author}
  {\bibfnamefont {Y.}~\bibnamefont {{Deng}}}, \bibinfo {author} {\bibfnamefont
  {S.}~\bibnamefont {{Chen}}}, \bibinfo {author} {\bibfnamefont {X.-J.}\
  \bibnamefont {{Liu}}}, \ and\ \bibinfo {author} {\bibfnamefont {J.-W.}\
  \bibnamefont {{Pan}}},\ }\href@noop {} {\bibinfo  {journal} {arXiv:1511.08170
  (2015)}\ }\BibitemShut {NoStop}%
\bibitem [{\citenamefont {Zhang}\ \emph {et~al.}(2013)\citenamefont {Zhang},
  \citenamefont {Chen},\ and\ \citenamefont {Zhang}}]{Zhang2013}%
  \BibitemOpen
\bibfield  {journal} {  }\bibfield  {author} {\bibinfo {author} {\bibfnamefont
  {Y.}~\bibnamefont {Zhang}}, \bibinfo {author} {\bibfnamefont
  {G.}~\bibnamefont {Chen}}, \ and\ \bibinfo {author} {\bibfnamefont
  {C.}~\bibnamefont {Zhang}},\ }\href@noop {} {\bibfield  {journal} {\bibinfo
  {journal} {Scientific reports}\ }\textbf {\bibinfo {volume} {3}} (\bibinfo
  {year} {2013})}\BibitemShut {NoStop}%
\bibitem [{\citenamefont {Jim\'enez-Garc\'{i}a}\ \emph
  {et~al.}(2015)\citenamefont {Jim\'enez-Garc\'{i}a}, \citenamefont {LeBlanc},
  \citenamefont {Williams}, \citenamefont {Beeler}, \citenamefont {Qu},
  \citenamefont {Gong}, \citenamefont {Zhang},\ and\ \citenamefont
  {Spielman}}]{PhysRevLett.114.125301}%
  \BibitemOpen
  \bibfield  {author} {\bibinfo {author} {\bibfnamefont {K.}~\bibnamefont
  {Jim\'enez-Garc\'{i}a}}, \bibinfo {author} {\bibfnamefont {L.~J.}\
  \bibnamefont {LeBlanc}}, \bibinfo {author} {\bibfnamefont {R.~A.}\
  \bibnamefont {Williams}}, \bibinfo {author} {\bibfnamefont {M.~C.}\
  \bibnamefont {Beeler}}, \bibinfo {author} {\bibfnamefont {C.}~\bibnamefont
  {Qu}}, \bibinfo {author} {\bibfnamefont {M.}~\bibnamefont {Gong}}, \bibinfo
  {author} {\bibfnamefont {C.}~\bibnamefont {Zhang}}, \ and\ \bibinfo {author}
  {\bibfnamefont {I.~B.}\ \bibnamefont {Spielman}},\ }\href {\doibase
  10.1103/PhysRevLett.114.125301} {\bibfield  {journal} {\bibinfo  {journal}
  {Phys. Rev. Lett.}\ }\textbf {\bibinfo {volume} {114}},\ \bibinfo {pages}
  {125301} (\bibinfo {year} {2015})}\BibitemShut {NoStop}%
\bibitem [{\citenamefont {Xu}\ \emph {et~al.}(2013)\citenamefont {Xu},
  \citenamefont {You},\ and\ \citenamefont {Ueda}}]{Xu2013}%
  \BibitemOpen
  \bibfield  {author} {\bibinfo {author} {\bibfnamefont {Z.-F.}\ \bibnamefont
  {Xu}}, \bibinfo {author} {\bibfnamefont {L.}~\bibnamefont {You}}, \ and\
  \bibinfo {author} {\bibfnamefont {M.}~\bibnamefont {Ueda}},\ }\href {\doibase
  10.1103/PhysRevA.87.063634} {\bibfield  {journal} {\bibinfo  {journal} {Phys.
  Rev. A}\ }\textbf {\bibinfo {volume} {87}},\ \bibinfo {pages} {063634}
  (\bibinfo {year} {2013})}\BibitemShut {NoStop}%
\bibitem [{\citenamefont {Anderson}\ \emph {et~al.}(2013)\citenamefont
  {Anderson}, \citenamefont {Spielman},\ and\ \citenamefont
  {Juzeliunas}}]{Anderson2013}%
  \BibitemOpen
  \bibfield  {author} {\bibinfo {author} {\bibfnamefont {B.~M.}\ \bibnamefont
  {Anderson}}, \bibinfo {author} {\bibfnamefont {I.~B.}\ \bibnamefont
  {Spielman}}, \ and\ \bibinfo {author} {\bibfnamefont {G.}~\bibnamefont
  {Juzeliunas}},\ }\href {\doibase 10.1103/PhysRevLett.111.125301} {\bibfield
  {journal} {\bibinfo  {journal} {Phys. Rev. Lett.}\ }\textbf {\bibinfo
  {volume} {111}},\ \bibinfo {pages} {125301} (\bibinfo {year}
  {2013})}\BibitemShut {NoStop}%
\bibitem [{\citenamefont {Goldman}\ and\ \citenamefont
  {Dalibard}(2014)}]{Goldman2014}%
  \BibitemOpen
  \bibfield  {author} {\bibinfo {author} {\bibfnamefont {N.}~\bibnamefont
  {Goldman}}\ and\ \bibinfo {author} {\bibfnamefont {J.}~\bibnamefont
  {Dalibard}},\ }\href {\doibase 10.1103/PhysRevX.4.031027} {\bibfield
  {journal} {\bibinfo  {journal} {Phys. Rev. X}\ }\textbf {\bibinfo {volume}
  {4}},\ \bibinfo {pages} {031027} (\bibinfo {year} {2014})}\BibitemShut
  {NoStop}%
\bibitem [{\citenamefont {Struck}\ \emph {et~al.}(2014)\citenamefont {Struck},
  \citenamefont {Simonet},\ and\ \citenamefont {Sengstock}}]{Struck2014}%
  \BibitemOpen
  \bibfield  {author} {\bibinfo {author} {\bibfnamefont {J.}~\bibnamefont
  {Struck}}, \bibinfo {author} {\bibfnamefont {J.}~\bibnamefont {Simonet}}, \
  and\ \bibinfo {author} {\bibfnamefont {K.}~\bibnamefont {Sengstock}},\ }\href
  {\doibase 10.1103/PhysRevA.90.031601} {\bibfield  {journal} {\bibinfo
  {journal} {Phys. Rev. A}\ }\textbf {\bibinfo {volume} {90}},\ \bibinfo
  {pages} {031601} (\bibinfo {year} {2014})}\BibitemShut {NoStop}%
\bibitem [{\citenamefont {{Luo}}\ \emph {et~al.}(2016)\citenamefont {{Luo}},
  \citenamefont {{Wu}}, \citenamefont {{Chen}}, \citenamefont {{Guan}},
  \citenamefont {{Gao}}, \citenamefont {{Xu}}, \citenamefont {{You}},\ and\
  \citenamefont {{Wang}}}]{LuoSR2016}%
  \BibitemOpen
  \bibfield  {author} {\bibinfo {author} {\bibfnamefont {X.}~\bibnamefont
  {{Luo}}}, \bibinfo {author} {\bibfnamefont {L.}~\bibnamefont {{Wu}}},
  \bibinfo {author} {\bibfnamefont {J.}~\bibnamefont {{Chen}}}, \bibinfo
  {author} {\bibfnamefont {Q.}~\bibnamefont {{Guan}}}, \bibinfo {author}
  {\bibfnamefont {K.}~\bibnamefont {{Gao}}}, \bibinfo {author} {\bibfnamefont
  {Z.-F.}\ \bibnamefont {{Xu}}}, \bibinfo {author} {\bibfnamefont
  {L.}~\bibnamefont {{You}}}, \ and\ \bibinfo {author} {\bibfnamefont
  {R.}~\bibnamefont {{Wang}}},\ }\href {\doibase 10.1038/srep18983} {\bibfield
  {journal} {\bibinfo  {journal} {Scientific Reports}\ }\textbf {\bibinfo
  {volume} {6}},\ \bibinfo {eid} {18983} (\bibinfo {year} {2016})}\BibitemShut
  {NoStop}%
\bibitem [{\citenamefont {Jotzu}\ \emph {et~al.}(2015)\citenamefont {Jotzu},
  \citenamefont {Messer}, \citenamefont {G\"org}, \citenamefont {Greif},
  \citenamefont {Desbuquois},\ and\ \citenamefont {Esslinger}}]{JotzuPRL2015}%
  \BibitemOpen
  \bibfield  {author} {\bibinfo {author} {\bibfnamefont {G.}~\bibnamefont
  {Jotzu}}, \bibinfo {author} {\bibfnamefont {M.}~\bibnamefont {Messer}},
  \bibinfo {author} {\bibfnamefont {F.}~\bibnamefont {G\"org}}, \bibinfo
  {author} {\bibfnamefont {D.}~\bibnamefont {Greif}}, \bibinfo {author}
  {\bibfnamefont {R.}~\bibnamefont {Desbuquois}}, \ and\ \bibinfo {author}
  {\bibfnamefont {T.}~\bibnamefont {Esslinger}},\ }\href {\doibase
  10.1103/PhysRevLett.115.073002} {\bibfield  {journal} {\bibinfo  {journal}
  {Phys. Rev. Lett.}\ }\textbf {\bibinfo {volume} {115}},\ \bibinfo {pages}
  {073002} (\bibinfo {year} {2015})}\BibitemShut {NoStop}%
\bibitem [{\citenamefont {Yu}\ \emph {et~al.}(2016)\citenamefont {Yu},
  \citenamefont {Xu}, \citenamefont {L\"u},\ and\ \citenamefont
  {You}}]{You2016PhysRevLett.116.143003}%
  \BibitemOpen
  \bibfield  {author} {\bibinfo {author} {\bibfnamefont {J.}~\bibnamefont
  {Yu}}, \bibinfo {author} {\bibfnamefont {Z.-F.}\ \bibnamefont {Xu}}, \bibinfo
  {author} {\bibfnamefont {R.}~\bibnamefont {L\"u}}, \ and\ \bibinfo {author}
  {\bibfnamefont {L.}~\bibnamefont {You}},\ }\href {\doibase
  10.1103/PhysRevLett.116.143003} {\bibfield  {journal} {\bibinfo  {journal}
  {Phys. Rev. Lett.}\ }\textbf {\bibinfo {volume} {116}},\ \bibinfo {pages}
  {143003} (\bibinfo {year} {2016})}\BibitemShut {NoStop}%
\end{thebibliography}


\end{document}